\documentclass[prl,onecolumn,floats,showpacs,superscriptaddress]{revtex4}
\usepackage{epsfig}
\usepackage{amsmath,amssymb,amsthm}
\usepackage{graphicx}
\usepackage{psfrag}
\usepackage{bm}
\newcommand{\lam}{\lambda}
\newcommand{\sig}{\sigma}
\newcommand{\al}{\alpha}
\newcommand{\LN}{L'}

\begin{document}
\title{Optimal network topologies: \\
Expanders, Cages, Ramanujan graphs, Entangled networks and all that.}

\author{Luca Donetti}
\affiliation{
Departamento de Electr\'onica and \\
Instituto de F{\'\i}sica
Te{\'o}rica y Computacional Carlos I, Facultad de Ciencias, Universidad de
Granada, 18071 Granada, Spain}

\author{Franco Neri}

\affiliation{Dipartimento di Fisica, Universit\`a di  Parma,
Parco Area delle Scienze 7/A, 43100 Parma, Italy}

\author{Miguel A. Mu{\~n}oz}

\affiliation{Departamento de Electromagnetismo y F{\'\i}sica de la Materia and \\
 Instituto de F{\'\i}sica Te{\'o}rica y Computacional Carlos I,
 Facultad de Ciencias, Universidad de Granada, 18071 Granada, Spain}

\date{\today} 
\begin{abstract} 
We report on some recent developments in the search for optimal
network topologies. First we review some basic concepts on spectral
graph theory, including adjacency and Laplacian matrices, and paying
special attention to the topological implications of having large
spectral gaps. We also introduce related concepts as ``expanders'',
Ramanujan, and Cage graphs. Afterwards, we discuss two different
dynamical features of networks: synchronizability and flow of random
walkers and so that they are optimized if the corresponding Laplacian
matrix have a large spectral gap. From this, we show, by developing a
numerical optimization algorithm that maximum
synchronizability and fast random walk spreading are obtained for a
particular type of extremely homogeneous regular networks, with long
loops and poor modular structure, that we call {\it entangled
networks}. These turn out to be related to Ramanujan and Cage
graphs. We argue also that these graphs are very good finite-size
approximations to Bethe lattices, and provide almost or almost optimal
solutions to many other problems as, for instance, searchability in
the presence of congestion or performance of neural networks. Finally,
we study how these results are modified when studying dynamical
processes controlled by a {\it normalized} (weighted and directed)
dynamics; much more heterogeneous graphs are optimal in this case.
Finally, a critical discussion of the limitations and possible
extensions of this work is presented.
\end{abstract} 


\maketitle

\section{I. Introduction}

Imagine you are to design a network, be it a local computer network,
an infrastructure or communication network, an artificial neural
network, or whatever analogous example you can think of. Suppose also
that you have some restrictions to do so. First of all, the number of
nodes and the total amount of links are both fixed and, second, nodes
should be connected using the available links in such a way that the
resulting network topology is {\it optimal} in some sense. Of course,
the meaning of the word ``optimal'' depends on the task to be
performed by the network, or in other words, depends on the nature of
the dynamical process to be built on top of it.

Along this paper we study optimization of dynamical processes in
networks, as exemplified by the following three cases.

Suppose we are constructing an artificial neural network, so at each
node we locate a neuron (i.e. an oscillator) having some unspecified
dynamical properties. In order to enhance the neural net performance
we want the oscillators to be easily synchronizable
\cite{synchro}, i.e. to be able to reach as easily as possible a state
in which them all, or at least a large fraction of them, fire
(oscillate) at unison \cite{Torres,BJK,GG}, and that state to be as
robust as possible.

As a second example, imagine having a communication or technological
network and wanting, for the sake of efficiency, any node to be
``nearby'' any other one. The simplest strategy to do so, is by
constructing a star-like topology with a central node (or ``hub'' in
the network jargon) directly connected to all the rest. In this way,
any node is reachable from any other within two steps at most.
However, in cases of intense traffic flow, this might not be very
efficient, especially if the central hub gets overburdened or
congestioned. Also, the star-like topology is very fragile to sabotage
to the central node. So one could wonder what is the optimal topology
avoiding the use of a privileged central hub
\cite{Catalans,Congestion,Maritan}.

For a third example, let us consider information packets traveling in
a network in such a way that they disperse jumping randomly between
contiguous nodes and let us require an optimum flow of
information. For that, we define an ensemble of random walkers
diffusing on the net, and impose the averaged first-passage time to be
minimized. Which is the optimal structure? Analogously, other
random-walk properties as the mixing rate \cite{Lovasz}, the mean
transit time or the mean return time \cite{dani} could be minimized.

In all these situations the problem to be solved is very similar:
finding an optimal topology under some constraints. These and similar
problems have been addressed in the literature, especially in the
context of Computer Science and, more recently, in the emerging field
of Complex Networks
\cite{Strogatz,Laszlo,Newman,Porto,AleRomu,Boccaletti}.

In this paper, we illustrate that the answer to these problems may
have a large degree of universality in the sense that, even if the
optimal topologies in each case are not ``exactly'' identical, they
may share some common features that we review here.  To do so, we
translate these problems into the one of optimizing some invariant
property of a matrix encoding the network topology. In particular, we
can use adjacency, Laplacian, and normalized-Laplacian matrices
depending on the particular problem under scrutiny. Spectral analysis
techniques are employed to explore in a systematic way network
topologies and to look for optimal solutions in each specific problem.
In this context, we introduce and characterize what we call {\it
entangled networks}: a family of nets with very homogeneous properties
and a extremely intertwined or entangled structure which are optimal
or almost optimal from different points of view.

Along this work, we focus mainly on un-weighted, un-directed,
networks, not embedded in a geography or physical coordinates, and
characterized by identical nodes. Changing any of these restrictions
may alter the nature of the emerging optimal topologies as will be
briefly discussed at the end of the paper.

The paper is structured as follows. In {\bf section II} we introduce
topological matrices and other basic elements of spectral graph
theory, as for instance the ``spectral gap''.  In {\bf section III} we
discuss the topological implications of having a large spectral gap
and illustrate the connections of this with topological optimization
problems: graphs with large spectral gaps are optimal (or almost
optimal) for some dynamical processes defined on networks. In {\bf
section IV} we present an explicit construction of optimal solutions,
called by graph theorists ``Ramanujan'' graphs; such explicit
constructions cannot be achieved for any number of nodes and
links. Owing to this, in {\bf section V} we discuss how to construct
optimal graphs in general (for any given number of nodes and edges) by
employing a recently introduced computational optimization method,
enabling us to construct the so called ``entangled networks''. We
compare entangled networks with Ramanujan graphs in the cases when
these last can be explicitly constructed, as well as with ``cage
graphs'', another useful concept in graph theory.  In {\bf section VI}
we enumerate some network design problems where our results are
relevant.  In {\bf section VII} we briefly elaborate on the connection
of Ramanujan and entangled networks in the limit of large sizes with
the Bethe lattice. In {\bf section VIII} we discuss how the previous
results are affected by analyzing dynamical processes controlled by
the normalized Laplacian rather than by the Laplacian. Finally, in
{\bf section IX} a critical discussion of our main results,
conclusions, and future perspectives is presented.


\section{II. Elements of spectral graph theory: spectral gaps}

For the sake of self-consistency and to fix notation we start by
revising some basic concepts in graph theory. 

Given a network (or graph), it is possible to define the corresponding
{\it adjacency matrix}, $A$, whose elements $a_{ij}$ are equal to $1$
if a link between nodes $i$ and $j$ exists and $0$ otherwise.  A
related matrix is the {\it Laplacian}, $L$, which takes values $-1$
for pairs of connected vertices and $k_i$ (the degree of the
corresponding node $i$) in diagonal sites. Obviously, $L = K - A$,
where $K$ is the diagonal connectivity (or degree) matrix.  If the
graph is undirected both $A$ and $L$ are symmetric matrices. It is
easy to see that $\lambda_1=0$ is a trivial eigenvalue of $L$ with
eigenvector $(1,1,\ldots)$ and that the eigenvalues $\lambda_i$
satisfy
\begin{equation*}
  0 = \lam_1 \le \lam_2 \le \ldots \le \lam_N \le 2k_{\text{max}} ,
\end{equation*}
where $k_{\text{max}}$ is the largest degree in the graph. The proofs
of these and other elementary spectral graph properties can be found,
for instance, in \cite{Chung,Bollobas}.  Finally, a {\it normalized
Laplacian matrix}, $\LN=K^{-1}L$, can be defined; each row is equal to
the analogous row in $L$ divided by the degree of the corresponding
node. This is useful to describe dynamical processes in which the
total effect of neighbors is equally normalized for all sites, while
in the absence of such a normalization factor sites with higher
connectivity are more strongly coupled to the others than loosely
connected ones.  In this case we have $0\le\lambda'_i\le2$ for
$i=1\ldots N$, where the $\lambda'_i$ denote the eigenvalues of $\LN$.
Clearly, for {\it regular} graphs, that is, graphs where all nodes
have the same degree $k$, the two matrices $L$ and $\LN$ differ by a
multiple of the identity matrix and the eigenvalues are related by
$\lambda_i=k \lambda'_i$.

Following the literature, here we consider mainly (but not only) the
Laplacian spectrum, but depending on the dynamical process under
consideration this choice will be changed.

In order to give a first taste on the topological significance of
spectral properties, let us consider a network perfectly separated
into a number of independent subsets or {\it communities} having only
intra-subset links but not inter-subset connections. Obviously, its
Laplacian matrix (either the non normalized or the normalized one) is
block-diagonal. Each sub-graph has its own associated sub-matrix, and
therefore $0$ is an eigenvalue of any of them. In this way, the
degeneration of the lowest (trivial) eigenvalue of $L$ (or of $\LN$)
coincides with the number of disconnected subgraphs within a given
network. 
For each of the separated components, the corresponding eigenvector
has constant components within each subgraph.

On the other hand, if the subgraphs are not perfectly separated, but a
small number of inter-subset links exists, then the degeneracy will be
broken, and eigenvalues and eigenvectors will be slightly
perturbed. In particular relatively small eigenvalues will appear, and
their corresponding eigenvectors will take ``almost constant'' values
within each subgraph. For instance, if the number of subsets is $2$,
spectral methods have been profusely used for graph bipartitioning or
graph bisecting \cite{GN} as follows. One looks for the smallest
non-trivial eigenvalue; its corresponding eigenvector should have
almost constant components within each of the two subgroups, providing
us with a bipartitioning criterion. Therefore a graph with a ``small''
first non-trivial Laplacian eigenvalue, $\lambda_2$, customarily
called {\it spectral gap} (or also {\it algebraic connectivity}) has a
relatively clean bisection. In other words, the smaller the spectral
gap the smaller the relative number of edges required to be cut-away
to generate a bipartition. Conversely a large ``spectral gap''
characterizes non-structured networks, with poor modular structure, in
which a clearcut separation into subgraphs is not inherent.  This is a
crucial topological implication of spectral gaps.

A closely related concept is the {\it expansion property}. To
introduce it, let us consider a generic graph, $X$, and define the
{\it Cheeger or isoperimetric constant} as follows. First, one
considers all the possible subdivisions of the graph in two disjoint
subsets of vertices: $A$ and its corresponding complement $B$. The
Cheeger constant, $h(X)$, is defined as the minimum value over all
possible partitions of the number of edges connecting $A$ with $B$
divided by the number of sites in the smallest of the two subsets. The
Cheeger constant is larger than $0$ if and only if the graph is
connected, and is ``large'' if any possible subdivision has many links
between the two corresponding subsets. Hence, for graphs with poor
community structure, the Cheeger constant is large.

In the graph-theory literature the concept of {\it expander} is
introduced within this context \cite{Sarnak,Valette}: a family of
expanders is a family of {\it regular graphs} with degree $k$ and $N$
nodes, such that for $N \rightarrow \infty$ the Cheeger constant $h$
is always larger than a given positive number $\epsilon$. Note that
the word ``expansion'', refers to the fact that the topology of the
network-connections is such that any set of vertices connects in a
robust way (``expands through'') all nodes, even if the graph is
sparse. Obviously, for this to happen, $k$ should be larger or equal
than $3$, as for $k=2$ all graphs are linear chains and can be,
therefore, cut in two subgraphs by removing just one edge (arbitrarily
small Cheeger constant for large system sizes).

The Cheeger constant and the expansion property can be formally
related to the spectral gap by the following inequalities \cite{AB}:
\begin{equation}
{\lam_2 \over 2} \leq h(X) \leq \sqrt{2k\lam_2}.
\label{inequality}
\end{equation}
Therefore, {\it a family of regular graphs is an expanding family if
  and only if it has a lower bound for the spectral gap, and the
  larger the bound the better the expansion}.  Among the applications
of expander graphs are the design of efficient communication networks,
construction of error-correcting codes with very efficient encoding
and decoding algorithms, de-randomization of random algorithms, and
analysis of algorithms in computational group theory \cite{Sarnak}.

As a consequence of this, expansion properties are enhanced upon
increasing the spectral gap, but spectral gaps cannot be as large as
wanted, especially for large graph sizes. In the case of $k$-regular
graphs (the ones for which expanders have been defined) there is an
asymptotic upper bound for it derived by Alon and Boppana \cite{AB},
given by the spectral gap of the infinite regular tree of degree $k$
(called Bethe tree or Bethe lattice):
\begin{equation}
\lim_{N \rightarrow \infty} \lam_2 \leq k - 2 \sqrt{k-1}.
\label{inequality2}
\end{equation} 

Whenever a graph has $ \lam_2 \geq k - 2 \sqrt{k-1}$ (i.e. when the
spectral gap is larger than its asymptotic upper bound), it is called
a {\it Ramanujan graph}. Therefore, Ramanujan graphs are very good
expanders.  In one of the following sections we will tackle the
problem of explicitly constructing Ramanujan graphs by following the
recently developed mathematical literature on this subject.

So far we have related the spectral graph to the ``compactness'' of a
given graph; a large gap characterizes networks with poor modular
structure, in which it is difficult to isolate sub-sets of sites
poorly connected with the rest of the graph or, in other words, large
gaps characterize expanders. In the following section we shift our
attention from topology to dynamics and discuss some other interesting
problems which turn out to be directly related to large spectral gaps.


\section{III. Large spectral gaps and dynamical processes}

In this section we review two different dynamical processes defined on
the top of networks and discuss how some of their properties depend on
the underlying graph topology \cite{Entangled}.

\subsection{Synchronization of dynamical processes}

An aspect of complex networks that has generated a burst of activity
in the last few years, because of both its conceptual relevance and
its practical implications, is the study of {\it synchronizability} of
dynamical processes occurring at the nodes of a given network. How
does synchronizability depend upon network topology? Which type of
topology optimizes the stability of a globally synchronized
state\cite{synchro}? 

A first partial answer to this question was given in a seminal work by
Barahona and Pecora \cite{Pecora} who established the following
criterion to determine the stability of fully synchronized states on
networks. Consider a general dynamical process
\begin{equation}
\dot{x}_i = F(x_i) + \sig \sum_{j\in n.n. i} [ H(x_j) -H(x_i)]= 
F(x_i) - \sig \sum_j L_{ij} H(x_j),
\label{pecora}
\end{equation}
where $x_i$ with $i \in {1,2,... ,N}$ are dynamical variables, $F$ and
$H$ are an evolution and a coupling function respectively, and
$\sigma$ is a constant. A standard linear stability analysis can be
performed by i) expanding around a fully synchronized state
$x_1=x_2=\ldots=x_N=x^s$ with $x^s$ solution of $\dot{x^s} = F(x^s)$,
ii) diagonalizing $L$ to find its $N$ eigenvalues, and iii) writing
equations for the normal modes $y_i$ of perturbations
\begin{equation}
 \dot{y}_i =
\left[ F'(x^s) - \sig
\lam_i H'(x^s) \right] y_i,
\label{linear}
\end{equation}
all of them with the same form but different effective couplings
$\al=\sig \lam_i$. Barahona and Pecora noticed that the maximum
Lyapunov exponent for Eq. (\ref{linear}) is, in general, negative only
within a bounded interval $[\alpha_A,\alpha_B]$, and that it is a
decreasing (increasing) function below (above) (see fig. 1 in
\cite{Pecora}, and see also the related work by Wang and Chen
\cite{Wang} in which the case $[\alpha_A,\infty]$ is studied). Requiring all
effective couplings to lie within such an interval, $ \al_A < \sig
\lam_2 \le \ldots \le \sig \lam_N <
\al_B$, one concludes that a synchronized state is linearly stable if
and only if $ \frac{\lam_N}{\lam_2} < \frac{\alpha_B}{\alpha_A}$ for
the corresponding network. It is remarkable that the left hand side
depends only on the network topology while the right hand side depends
exclusively on the dynamics (through $F$ and $G$, and $x^s$).

As a conclusion, the interval in which the synchronized state is
stable is larger for smaller eigenratios $\lam_N/\lam_2$, and
therefore one concludes that {\it a network has a more robust
synchronized state if the ratio $Q = \lam_N/\lam_2$ is as small as
possible} \cite{Wang}. Also, as the range of variability of $\lam_N$
is limited (it is related to the maximum connectivity \cite{Mohar})
minimizing $Q$ gives very similar results to maximizing the
denominator $\lam_2$ in most cases. Indeed, as argued in \cite{Wang}
in cases where the maximum Lyapunov exponent is negative in an
un-bounded from above interval, the best synchronizability is obtained
by maximizing the spectral gap. 

It is straightforward to verify that for normalized dynamics, i.e.,
problems where a quotient $k_i$ appears in the coupling function in
Eq.(\ref{pecora}), the Laplacian eigenvalues have to be replaced by
the normalized-Laplacian ones, so the gap refers to $\lambda'_2$ and
$Q$ becomes $Q_{norm}= \lam'_N/\lam'_2$.

Summing up, {\it large spectral gaps favor stability of synchronized
states (synchronizability)}.

\subsection{Random walk properties}

One of the first and most studied models on graphs are random walks:
these are important both as a simple models of dispersion phenomena
and as a tool for exploring graph properties.

Indeed random walk properties are strictly related to the underlying
network structure and, in particular, to the eigenvalue spectrum of
the previously introduced matrices. At each step the transition
probability from vertex $i$ to vertex $j$ is trivially given by
$P_{ij} = {A_{ij} \over k_i}$. This defines a transition matrix $P$
for random walk dynamics, that can be written as $P= K^{-1} A =
K^{-1} (K - L) = I -\LN$ (where $I$ the identity matrix); therefore
the eigenvectors of $P$ and $\LN$ coincide and the eigenvalues are
linearly related.

In particular, the stationary probability distribution of the random
walk on a graph is given by the eigenvector corresponding to the
largest eigenvalue (1) of $P$, corresponding to $\lambda'=0$.  Then it
is easy to see that the convergence rate of a given initial
probability distribution towards its stationary distribution (also
called ``mixing rate'') is controlled by the second largest eigenvalue
of $P$, and therefore by the spectral gap of the normalized Laplacian
matrix: the larger $\lambda'_2$ the faster the decay
\cite{Nielsen,Lovasz}.

Another relevant property is the first passage time, $\tau_{i,j}$
between two sites $i$ and $j$, defined as the average time it takes
for a random walker to arrive for the first time to $j$ starting from
$i$. It can be expressed in terms of the eigenvectors ($u_k$) and
eigenvalues of the normalized Laplacian matrix as
\begin{equation*}
  \tau_{i,j} = 2 M \sum_{l=2...N} { \left(\frac{u_{l,i}}{\sqrt{k_i}}
      -\frac{u_{l,j}}{\sqrt{k_j}}\right)^2 \frac1{\lambda'_l} },
\end{equation*}
where $M$ is the number of link of the network (see \cite{Lovasz}).
For $k-$regular graphs the expression of its graph average, $\tau$,
can be simplified in the following way:
\begin{equation}
\tau = 2 k \frac{N}{N-1} \sum_{l=2...N} \frac1\lambda_l
\label{fp}
\end{equation}
A large $\lambda_2$ (which imply that also the following eigenvalues
cannot be small) gives an important contribution for keeping $\tau$ small.
Therefore large spectral gaps are associated with short first-passage times.

Finally, random walks move around quickly on graphs with large
spectral gap in the sense that they are very unlikely to stay long
within a given subset of vertices, $A$, unless its complementary
subgraph is very small \cite{Nielsen,Lovasz}.  This idea has been
exploited by Pons and Latapy \cite{PL} to detect communities
structures in networks by associating them to regions where random
walks remain trapped for some time. In conclusion: random walks escape
quickly from any subset $A$ if the spectral gap is large.

Summing up: {\it random walks move and disseminate fluently in large
spectral gap graphs}.  Related results, more details, and proofs of
these theorems can be found, for example in \cite{Nielsen,Lovasz}.


\section{IV. Explicit construction of Ramanujan Graphs}

The summary of the preceeding section is that if we are aimed at
designing network topologies with good synchronizability or
random-walk flow properties, we need criteria to construct graphs with
large spectral gaps.  On the other hand, as explained before,
Ramanujan graphs are optimal expanders, in the sense that a given
family of them, with growing $N$, will converge asymptotically from
above to the maximum possible value of the spectral gap.  Even if this does
not imply that for finite arbitrary values of $N$ they are optimal,
i.e. that they have the {\it largest} possible spectral gap, they
provide a very useful approach for the optimization of the spectral
gap problem.  Therefore, a good way to design optimal networks so is
to explicitly construct Ramanujan graphs by following the mathematical
literature on this respect  \cite{Valette,LPS,Margulis,Chiu}.

In this section we present a recipe for constructing explicitly
families of Ramanujan graphs while, in the following one, we will
construct large-spectral gap networks by employing a computational
optimization procedure. Readers not interested in the mathematical
constructions can safely skip this section.

\subsection{The recipe}

In recent years some explicit methods for the construction of
Ramanujan graphs have appear in the mathematical literature
\cite{LPS,Margulis}. We describe here only one of them.
While the proof that these graphs, constructed by Lubotzky, Phillips
and Sarnak \cite{LPS} (and independently by Margulis \cite{Margulis})
are Ramanujan graphs is a highly non-trivial one, their construction,
following the recipe we describe in what follows, is relatively simple
and can be implemented without too much effort.

Let us consider a given group $G$ and let $S \in G$ be a subset of
group elements not including the identity. The {\it Cayley graph}
associated with $G$ and $S$ is defined as the directed graph having
one vertex for each group element and directed edges connecting them
whenever one goes from one group element (vertex) to the other by
applying a group transformation in $S$. The absence of the identity in
$S$ guarantees that self-loops are absent. The Cayley graph depends on
the choice of the generating set $S$ and it is connected if and only
if $S$ generates $G$. Indeed, this will be the only case we consider
here.  Note that if for each element $s \in S$ its inverse $s^{-1}$
also belongs to $S$, then the associated Cayley graph becomes
undirected, which is the case in all what follows.

In the construction we discuss here, the group $G$ is given by
either by $PGL(2,Z/qZ)$ or by $PSL(2,Z/qZ)$. Consider the group of $2
\times 2$ matrices with elements in $Z/qZ$ (i.e. integer numbers
modulo $q$, where $q$ is an odd prime number); the elements of
$PGL(2,Z/qZ)$ ($PSL(2,Z/qZ)$) are the equivalence classes of matrices
with non-vanishing determinant (determinant equal to one) with respect
to multiplications by multiples of the identity matrix (i.e. to
matrices differing in multiples of the identity are considered to be
equal).

To specify the subset $S$ of Cayley graph generators, the integral
quaternions $H(Z) = \{ \alpha = a_0 + a_1 i + a_2 j + a_3 k: a_i \in Z
\} $ must be introduced. Quaternions can be casted in a matrix-like
form as \cite{Valette}:
\begin{equation}
 \left( \begin{array}{cc}
 a_0 + a_1 x +a_3 y & -a_1 y + a_2 + a_3 x \\
-a_1 y -a_2 + a_3 x & a_0 - a_1 x -a_3 y \end{array} \right)
\label{matrix}
\end{equation}
where $x,y$ are odd prime integers, satisfying $(x^2 + y^2 +1)\mod q =
0$. Writing quaternions in this way, it turns out that their algebra
coincides with the standard algebra of matrices and the quaternion
norm $|\al|^2 = a_0^2+a_1^2+a_2^2+a_3^2$ is equal to the corresponding
matrix determinant. Now another odd prime number $p$ has to be
chosen. It can be shown that there are $8 (p+1)$ elements in $H(Z)$
satisfying $|\al|^2=p$. If $p \mod 4 = 1$ then by taking $a_0$ odd and
positive the number of solutions is reduced to $p+1$, while if $p \mod
4 = 3$ then the number of solutions is reduced to $p+1$ by taking
$a_0$ even and requiring the first non-zero component to be positive.

If $p$ is a perfect square modulo $q$ such solutions can be mapped to
a set $S$ of $p+1$ matrices belonging $PSL(2,Z/qZ)$ (to guarantee that
all such matrices are distinct, one can take $q>2\sqrt{p}$). A
Ramanujan graph is then built as the Cayley graph with $G=PSL(2,Z/qZ)$
using the set $S$ of generators we just constructed; the number of
nodes is given by $N=q(q^2-1)/2$ (the number of elements of
$PSL(2,Z/qZ)$) and the degree of each node is $k=p+1$ (the elements of
$S$). If, otherwise, $p$ is not a perfect square modulo $q$ we map the
solutions to a set $S$ of $PGL(2,Z/qZ)$ matrices and a Ramanujan graph
is build as the Cayley graph with $G=PGL(2,Z/qZ)$ and the set $S$ just
constructed.  In this case $N=q(q^2-1)$ and $k=p+1$.

Putting all this together, Ramanujan graphs can be built by the
following procedure \cite{Code,Valette}:

\begin{itemize}
\item choose values of the odd prime numbers $q$ and $p$, with $q>2\sqrt{p}$;

\item associate the matrices of $PGL(2,Z/qZ)$ or $PSL(2,Z/qZ)$
  (depending on the relative values of $p$ and $q$) with graph nodes;

\item find the solutions of $a_0^2+a_1^2+a_2^2+a_3^2=p$, a solution of
  $(x^2 + y^2 +1)\mod q = 0$ and construct the matrices according to
  (\ref{matrix});

\item find the neighbors of each node by multiplying the corresponding
  matrix by the just constructed $p+1$ matrices, and represent them by
  edges.

\end{itemize}

In this way we construct regular graphs with a large spectral gap,
with some restrictions on the number of nodes $N$ and the degree $k$:
the degree can only be an odd prime number plus one, while the number
of nodes grows as the third power of $q$.

\subsection{Topological properties of the resulting graphs}

\begin{figure}
\centerline{\psfig{file=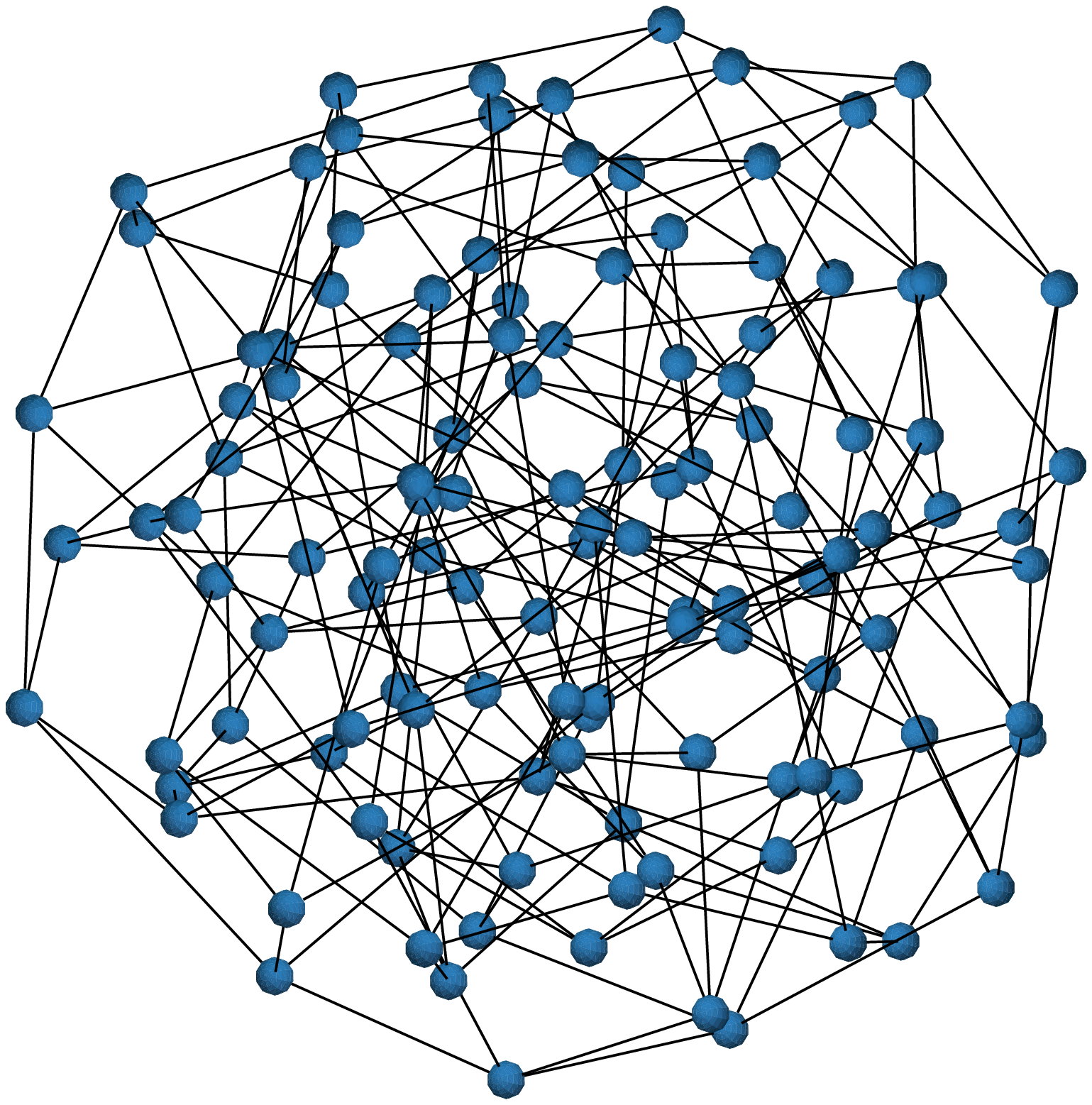,width=6.0cm}
\psfig{file=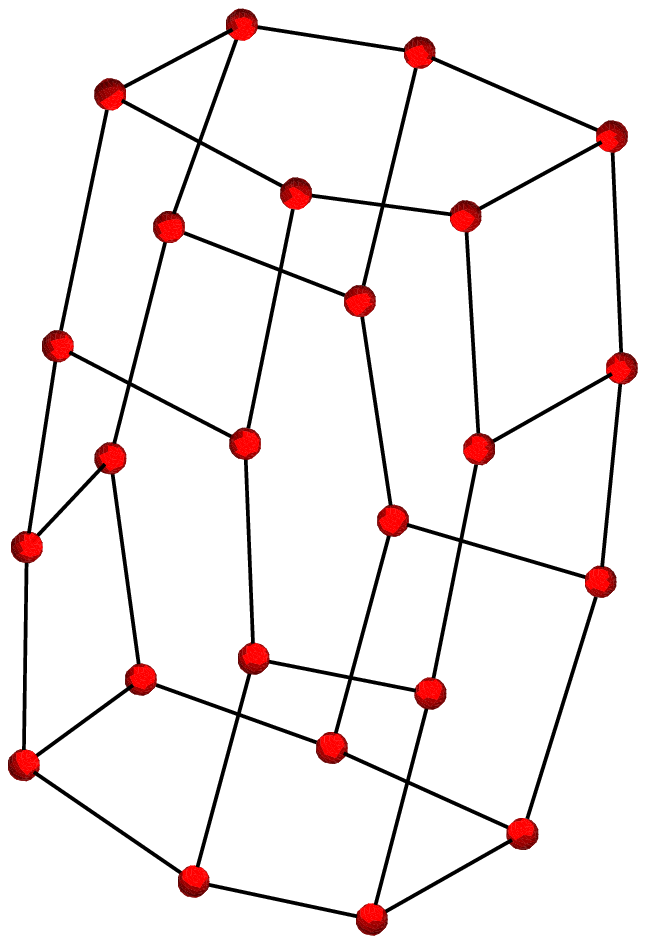,width=7.0cm}}
\caption{
Ramanujan graphs with (left) degree $k=4$ ($p=3$) and $120$ nodes
($q=5$), and (right) $k=3$ ($p=2$) and $24$ nodes ($q=3$).}
\label{Rama}
\end{figure}

First of all, every Cayley graph is by construction a regular graph,
and so are the Ramanujan graphs built following the above procedure.
Moreover, Lubotzky, Phillips and Sarnak \cite{LPS} proved that
these graphs have a large girth (the {\it girth}, $g$, of a graph is
the length of the shortest loop or cycle, if any), providing a lower
bound for the $g$ which grows logarithmically with $q$ (its exact form
is different in the $PSL$ and $PGL$ cases).  To gain some more
intuition about the properties of this family of graph, we plot in the
left of figure (\ref{Rama}) the smallest one, corresponding to $k=4$
($p=3$) and $120$ nodes ($q=5$).

Some remarks on its topological properties are in order.  The average
distance $3.714$ is relatively small (i.e. one reaches any of the
$120$ nodes starting from any arbitrary origin in less than four
steps, on average). The betweenness centrality \cite{betweenness} $
161.5$ is also relatively small, and takes the same value for all
nodes. The clustering coefficient vanishes, reflecting the absence or
short loops (triangles), and the minimum loop size is large, equal to
$6$, and identical for all the nodes.  In a nutshell: {\it the network
homogeneity is remarkable; all nodes look alike, forming rather
intricate, decentralized structure, with delta-peak distributed
topological properties.}

\subsection{Extension to $k=3$}

P. Chiu, extended the previous method to consider also graphs with
degree $3$ (i.e. $p=2$; note that the previous construction was
restricted to odd prime values of $p$). To do this, one just needs to
consider the following $3$ generators \cite{Chiu}:
\begin{equation}
\left( \begin{array}{cc}
1 & 0 \\ 0 & -1 \end{array} \right), \left( \begin{array}{cc}
2+\sqrt{-2} & \sqrt{-26} \\ \sqrt{-26} & 2-\sqrt{-2} \end{array}
\right), \left( \begin{array}{cc} 2-\sqrt{-2} & -\sqrt{-26} \\ -
\sqrt{-26} & 2+\sqrt{-2} \end{array} \right),
\label{matrix7}
\end{equation}
acting on the elements of $PGL(2,Z/qZ)$ or $PGL(2,Z/qZ)$ as before.

In the right part of figure (\ref{Rama}) we show the smallest
Ramanujan graph with degree $k=3$ ($p=2$) and $24$ nodes ($q=3$).
This is slightly less homogeneous than the one with $k=4$, but it is
as homogeneous as possible given the previous values of $N$ and
$k$. The average distance is also small in this case, $3.13$, the
betweenness is $24.50$ for all nodes, the clustering vanishes, and
the size of the minimum loop is $4$.  While for edges, the edge
betweenness \cite{betweenness} is $27.31$ for $12 $ edges while it is
$22.34$, for the remaining $24$ edges; indicating that the net is
not fully homogeneous in this case, but not far from homogeneous
either.

Summing up, the main topological features of these Ramanujan graphs is
that they are very homogeneous: properties as the average distance,
betweenness, minimum loop size, etc are very narrowly distributed
(they are delta peeks in many cases). Also, compared to generic random
regular graphs, their averaged distance is smaller and the
average minimum loop size is larger.  A main limitation of Ramanujan
graphs constructed in this way is that, for a fixed connectivity, the
possible network sizes are restricted to specific, fast growing
values.

\section{V. Construction of networks by computational optimization}

An alternative route to build up optimal networks with an arbitrary
number of nodes and an as-large-as-possible spectral gap or (almost
equivalently) an as-small-as-possible synchronizability ratio
$Q=\lambda_N/\lambda_2$ is by employing a computational optimization
process. Note, that enumerating all possible graphs with fixed $N$ and
$\langle k \rangle$, and looking explicitly for the minimum value of
$Q$ or largest gap, is a non-polynomial problem and, hence,
approximate optimization approaches are mandatory \cite{Entangled}.
In this section, we focus on optimizing the Laplacian spectral gap
while in section VIII we will tackle the normalized Laplacian case and
discuss the differences between the two of them.

\subsection{The algorithm}

The idea is to implement a modified {\it simulated annealing}
algorithm \cite{SA} which, starting from a random network with the
desired number of nodes $N$ and average connectivity-degree $ \langle
k \rangle$, and by performing successive rewirings, leads
progressively to networks with larger and larger spectral gaps or
smaller and smaller eigenratios $Q$. As the spectral gap will be very
large in the emerging networks, they will be typically Ramanujan
graphs (whenever they are regular).

The computational algorithm is as follows \cite{Entangled}. At each
step a number of rewiring trials is randomly extracted from an
exponential distribution. Each of them consists in removing a randomly
selected link, and introducing a new one joining two random nodes
(self-loops are not allowed). Attempted rewirings are (i) rejected if
the updated network is disconnected, and otherwise (ii) accepted if
$\delta Q= Q_{final}-Q_{initial} <0$, or (iii) accepted with
probability
\cite{Penna} $p = \min\left( 1, [1-(1-q) \delta Q/T]^{1/(1-q)}\right)$
(where $T$ is a temperature-like parameter) if $\delta Q \geq 0$. Note
that, in the limit $q \to 1$ we recover the usual Metropolis
algorithm. Instead we choose $q=-3$ which we have verified to give the
fastest convergence, although the output does not essentially depend
on this choice \cite{Penna}. The first $N$ rewirings are performed at
$T=\infty$. They are used to calculate a new $T$ such that the largest
$\delta Q$ among the first $N$ ones would be accepted with some large
probability: $T=(1-q)\cdot(\delta Q)_{max}$. Then $T$ is kept fixed
for $100N$ rewiring trials or $10N$ accepted ones, whichever occurs
first. Afterwards, $T$ is decreased by $10\%$ and the process iterated
until there is no change during $5$ successive temperature steps,
assuming that a (relative) minimum of $Q$ has been found. It is
noteworthy that most of these details can be changed without affecting
significatively the final results, while two algorithm drawbacks are
that the calculation of eigenvalues is slow and that the dynamic can
get trapped into ``metastable states'' (corresponding to local but not
global extreme values) as we illustrate in \cite{Entangled}.

\begin{figure}
\centerline{
\psfig{file=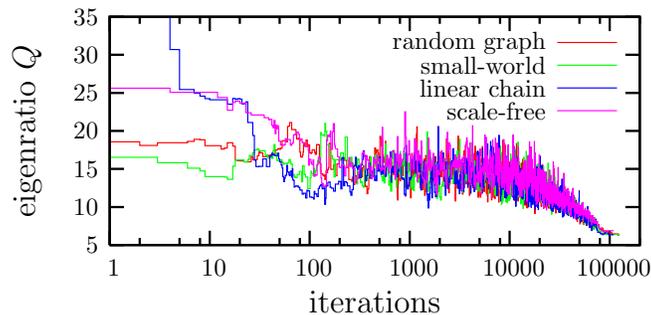,width=8.5cm}}
\caption{Eigenvalue ratio, $Q$ as a function of the number of algorithmic
iterations, starting from different initial conditions (random
network, small-world, linear chain, and scale free network) with
$N=50$, and $\langle k \rangle=4$. The algorithm leads to topologies
as the one depicted in figure 4.}
\label{Evolution}
\end{figure}

Running the algorithm for different initial networks, we find that
whenever $N$ is small enough (say $N \lesssim 30$), the
output-topology is unique in most of the runs, while some dispersion
in the outputs is generated for larger $N$ ($N=2000$ is the larger
size we have optimized). This is an evidence that the $Q$ absolute
minimum is not always found for large graphs, and that the evolving
network can remain trapped in metastable states. In any event, the
final values of $Q$ are very similar from run to run, starting with
different initial conditions, as shown in fig.~\ref{Evolution}, which
makes us confident that a reasonably good and robust approximation to
the optimal topology is typically obtained, though, strictly speaking,
we cannot guarantee that the optimal solution has been actually found,
especially for large values of $N$.

\subsection{Topological properties of the emerging network}
\begin{figure}
\centerline{
\psfig{file=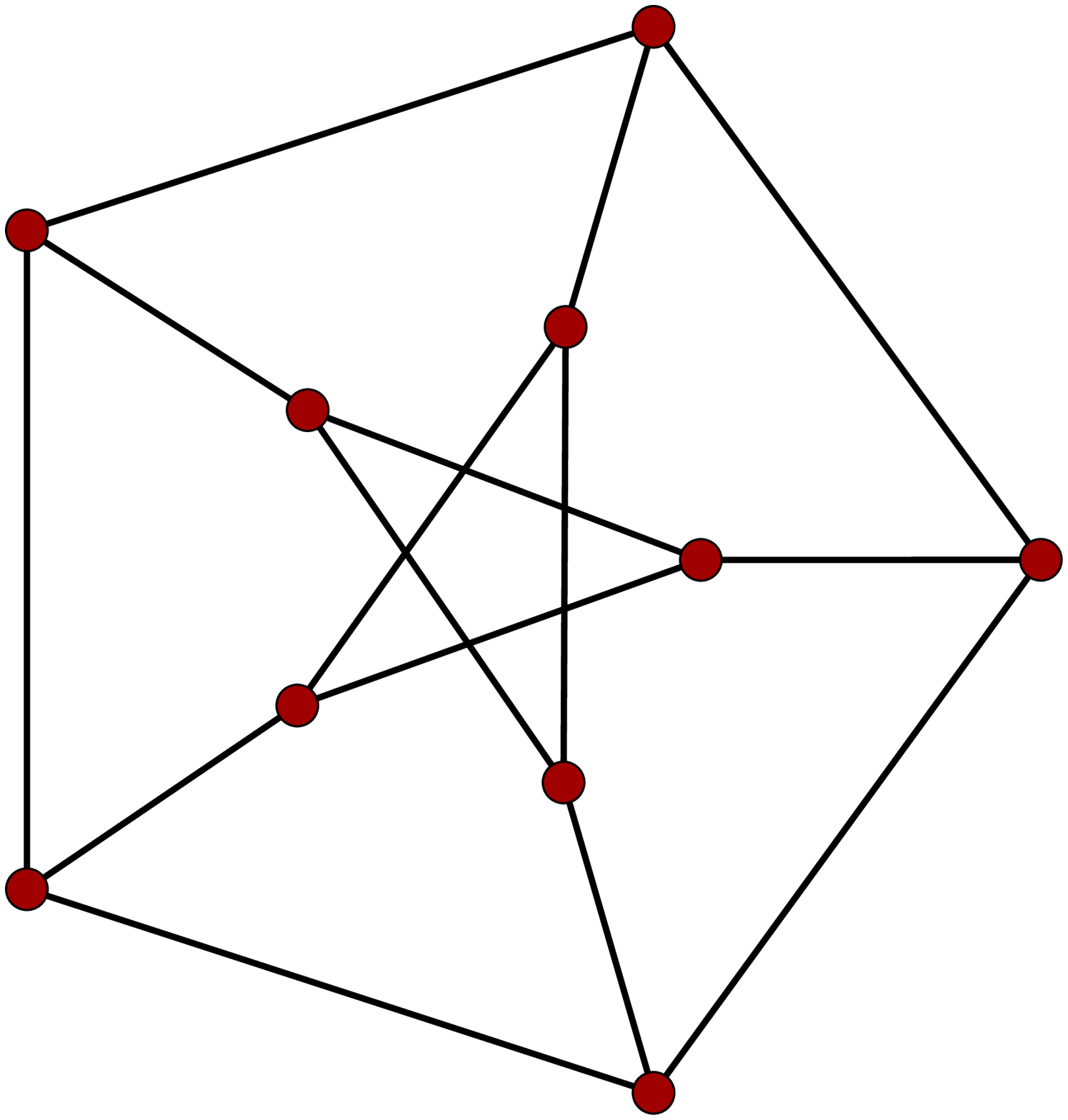,width=3.5cm}
\hspace{2cm}
\psfig{file=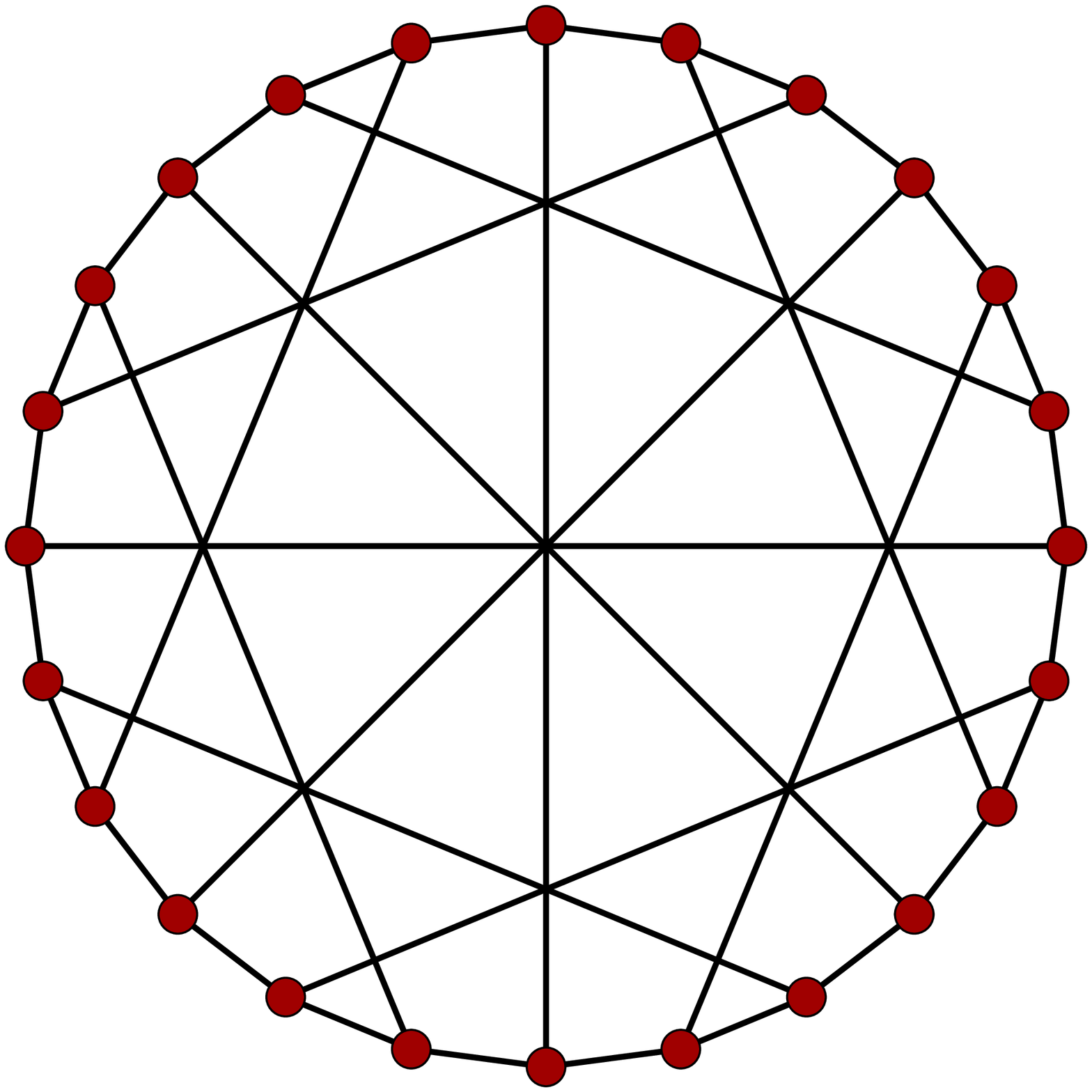,width=3.5cm}}
\caption{Optimal, entangled, networks obtained as output of the optimization
procedure for $k =3$ with $N=10$ and $N=24$. The left one is a
Petersen cage graph (k=3, girth=5). The one to the right is a McGee
cage graph (k=3, girth=7).}
\label{Cages}
\end{figure}

\begin{figure}
\centerline{
\psfig{file=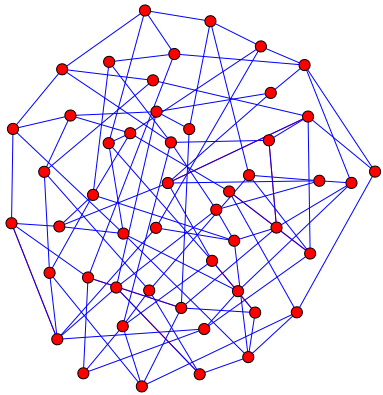,width=5.0cm}}
\caption{Optimal, entangled, network obtained as output of the optimization
procedure for $k=4$ with $N=50$.}
\label{Entangled}
\end{figure}

In figures (\ref{Cages}) and (\ref{Entangled}) we illustrate the
appearance of the networks emerging out of the optimization procedure,
that we call {\it entangled networks}, for different values of $N$ and
$\langle k \rangle$.
Remarkably, for some small values of $N$ and $k$ (see figure
(\ref{Cages})), it is possible to identify the resulting optimized
networks with well-known ones in graph theory: {\it cage graphs}
\cite{Cages,Bollobas}. 

\begin{figure}
\centerline{
\psfig{file=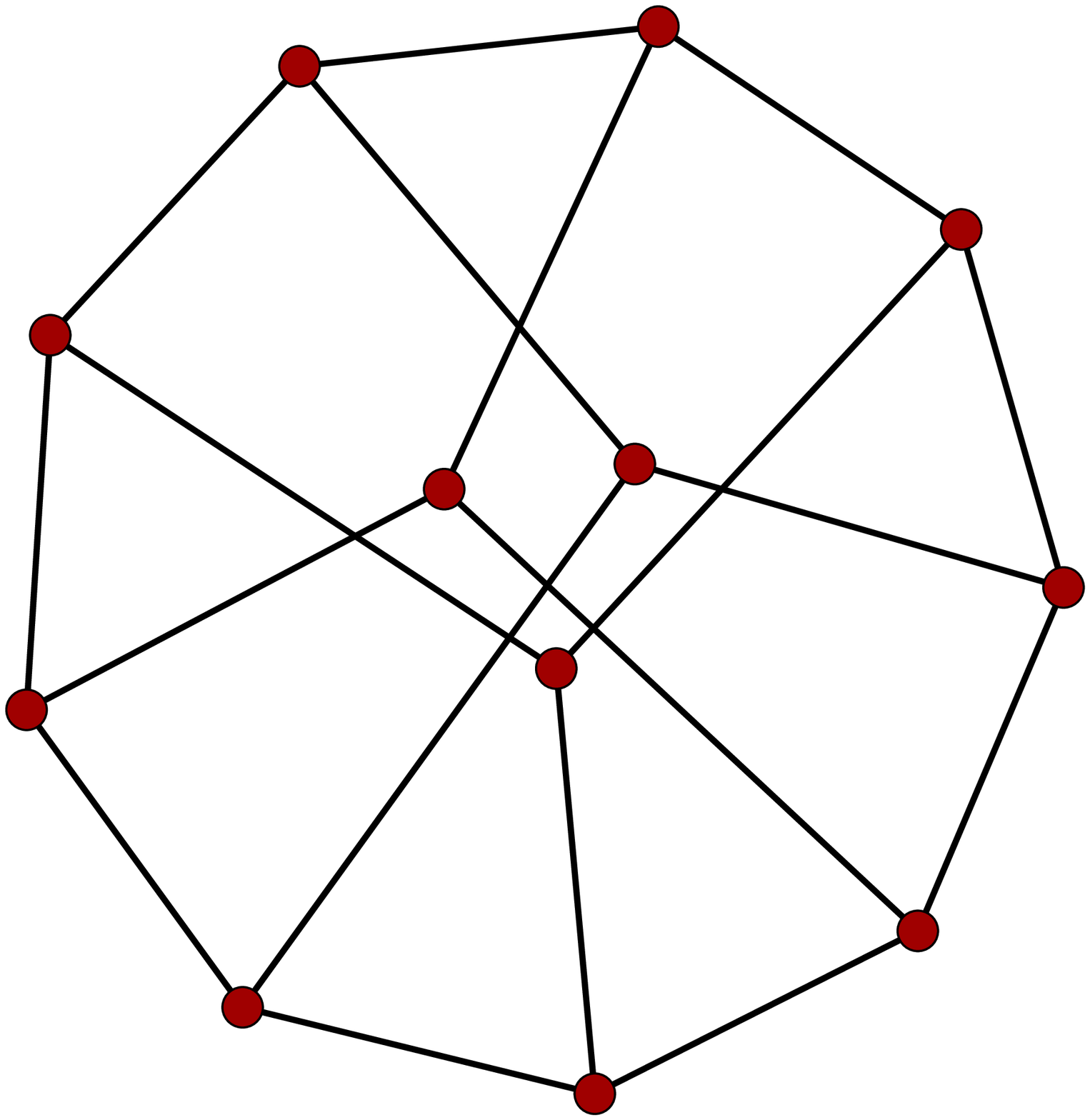,width=3.5cm}
\hspace{2cm}
\psfig{file=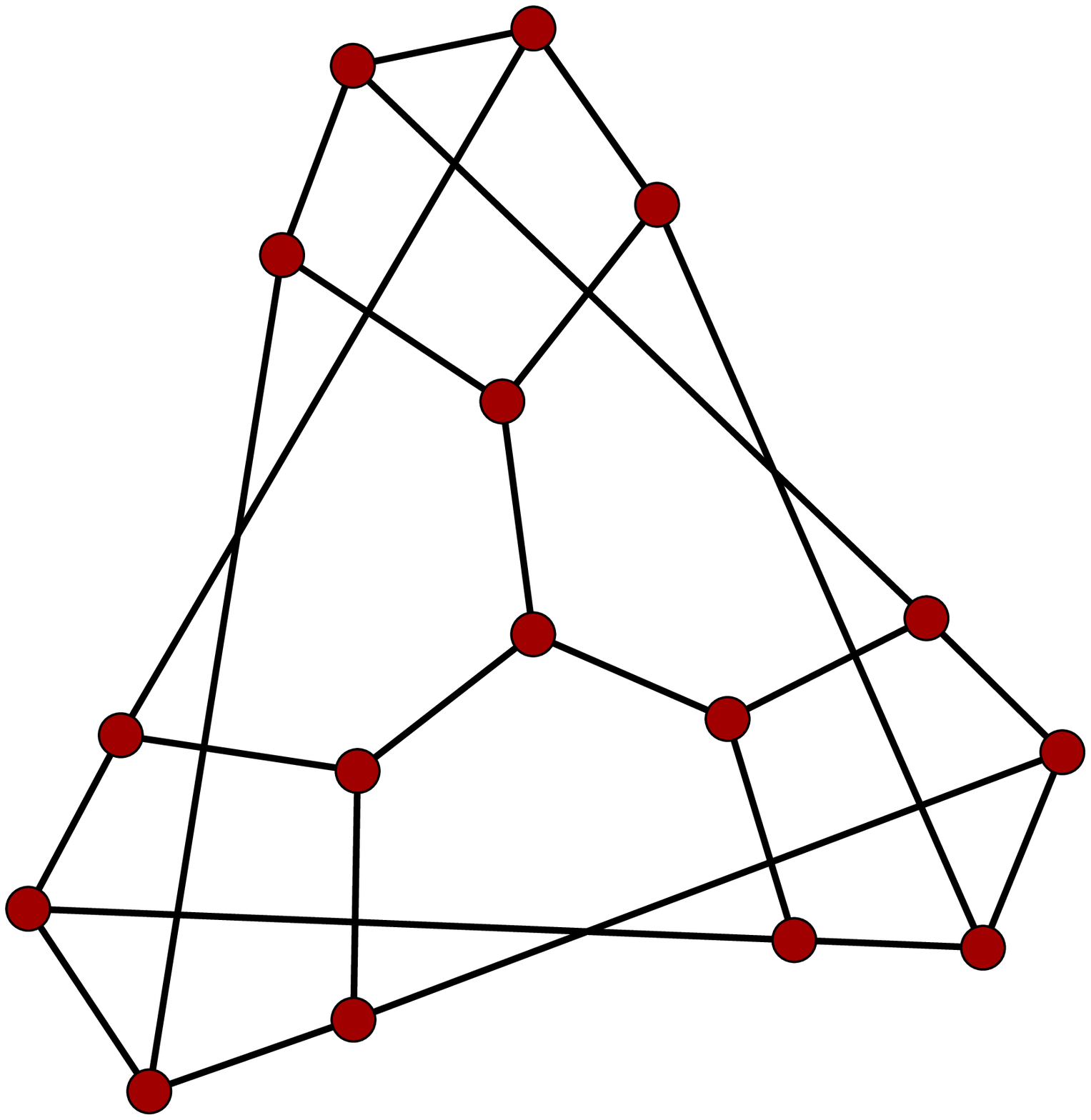,width=3.5cm}}
\caption{Optimal, entangled, networks obtained as output of the optimization
procedure for $k =3$ with $N=12$ and $N=16$. They do not correspond to
cage-graphs.}
\label{NoCages}
\end{figure}

A $(k,g)$-cage graph is a $k$-regular graph with girth $g$ having the
minimum possible number of nodes.  Cage graphs have a vast number of
applications in computer science and theoretical graph analysis
\cite{Cages,Bollobas}.  For $k=3$ and $N=10, 14$, and $24$,
respectively, the optimal nets found by the algorithm are cage-graphs
with girth $5$, $6$, and $7$ respectively (called $Petersen$,
$Heawood$ and $McGee$ graphs in the mathematical literature; see
fig.(\ref{Cages}) and also the nice picture gallery and mathematical
details in \cite{Cages}). We also recover other cage graphs for small
values of $N$ for $k=4$ and $k=5$.
For some other small values of $N$, cage graphs do not exist. For
example, for $k=3$ and girths, $3, 4, 5, 6$, and $7$ the Cage graphs have
the sizes: $N=4, 6, 10, 14$, and $24$ respectively
\cite{Cages}. Therefore, for sizes as $N=12$ or $N=16$ there is no
cage with $k=3$. In these cases, our optimization procedure leads to
graphs similar to cages (see figure
\ref{NoCages}) in which the shortest loops are as large as possible,
and all of them have very similar lengths. In this sense, our family
of optimal graphs provides us with topologies ``interpolating''
between well known cage graphs.

In the emerging entangled networks short loops are severely suppressed
as said before. This can be quantified by either the {\it girth} or
more accurately by the average length, $\langle \ell \rangle$, of the
shortest loop from each node. In particular, the clustering
coefficient (measuring the number of triangular loops in the net)
vanishes, as loops have typically more than three edges.

\begin{figure}
\centerline{
\psfig{file=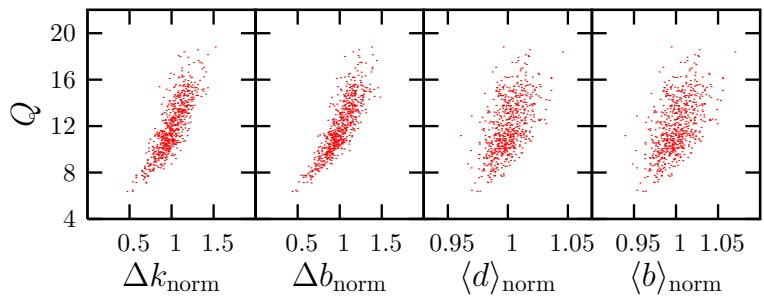,width=10.0cm}}
\caption{Relation between the ratio $Q$ and (i) node-degree standard
deviation, (ii) betweenness standard deviation, (iii) average
node-distance, and (iv) average betweenness. The subscript ``norm''
stands for normalization with respect to the respective mean-values,
centering all the measured quantities around $1$. They all decrease on
average as the optimization procedure goes.}
\label{evolution}
\end{figure}

To have a more precise characterization of the emerging topologies,
especially for larger graphs (see figure (\ref{Entangled})), we
monitored different topological properties during the optimization
process, as shown in figure (\ref{evolution}).
The node-degree standard deviation typically decreases as $Q$
decreases meaning that the optimal topology approaches a regular one.
The betweenness standard deviation also decreases with $Q$ implying
that optimal networks tend to be as homogeneous as possible, and that
all nodes play essentially the same role in flow properties.  Also,
the average node distance and average betweenness are progressively
diminished on average: nodes are progressively closer to each other
and the network becomes less centralized as the optimization procedure
runs. It has also been numerically verified that the average first
passage times of random walk is reduced during the network
optimization.

The fact that the betweenness distribution is very narrow, indicates
that (following an original idea by Girvan and Newman \cite{GN}) it is
very difficult to divide the network into communities. Indeed, in the
algorithm proposed in \cite{GN} to identify communities, links with
high betweenness centrality are progressively cut out; if all links
have the same centrality the method becomes useless and communities
are hardly distinguishable.

We have named the emerging structures {\it entangled networks}, to
account for their very intricate and interwoven topology, with
extremely high homogeneity in different topological properties, poor
modularity or community structure, and large loops.  In particular, as
the spectral gap is very large and they tend to be regular, entangled
nets are usually Ramanujan graphs.

\begin{figure}
\centerline{ 
\psfig{file=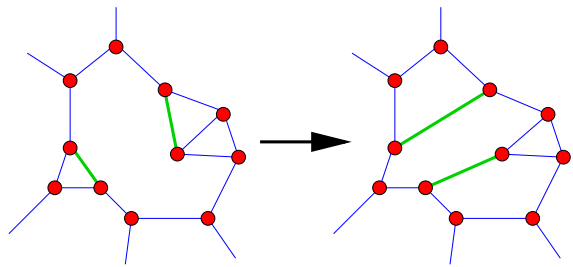,width=7.5cm}}
\caption{Rewiring attempt in the connectivity conserving
version of the algorithm.}
\label{rewiring}
\end{figure}

We have also exploited our understanding of the entangled-topology to
generate optimal networks more efficiently. In particular, given the
convergence in all the explored cases to almost regular networks, we
have constructed an improved version of the algorithm in which we
start with random regular nets and rewire edges in such a way that the
original degree distribution is preserved (indeed, most of the
entangled networks depicted in this section are obtained using this
improved algorithm). For this, links are selected by pairs, an origin
node is selected for each one and the two end nodes are exchanged as
shown in figure \ref{rewiring}. We have observed numerically that the
convergence towards optimal solutions is faster when constraining the
optimization to the regular networks and that for large values of $N$
the final outputs have smaller values of $Q$ than the original ones
(owing to the fact that it is less likely to fall into metastable
states).  The discussed topological traits remain unaffected.

\subsection{Comparing Cayley-graph Ramanujans with Entangled networks}

\subsubsection{Small sizes: Cage graphs}

If we are to design optimal networks with a small number of vertices
(say $N < 100$) then entangled networks (and hence, cage-graphs,
whenever they exist) are a better choice than Ramanujan nets based on
Cayley-graphs, as those constructed in the previous section. First,
these second do exist only for a few small values of $N$. Second,
because even when such Ramanujan graphs exist entangled network
outperform them, as they have a smaller value of $Q$ (larger spectral
gap) which has been optimized on-purpose.

\subsubsection{Large sizes}
For large values of $N$ (say $N >100$) entangled networks are
difficult to construct as the optimization algorithm becomes not
affordably time-consuming. Instead, Cayley-graph Ramanujan networks,
whenever they can be constructed, are a better choice. They are easy
and fast to construct by following the recipe described in a previous
section and they provide large spectral gaps, closer and closer to the
optimal value (asymptotic upper bound) as $N$ increases.

\section{VI. Related problems and similar topologies}

In this section we illustrate how entangled networks and Ramanujan
graphs play a crucial role in different contexts, and emerge as
optimal (or close to optimal) solutions for a number of network
optimization problems.

\subsection{Local search with congestion}

This is one of the examples illustrated in the introduction, and has
been recently tackled by Guimer\'a et al. \cite{Catalans}. By defining
an appropriate {\it search cost} function these authors explore which
is the ideal topology to optimize searchability and facilitate
communication processes.  They arrive at the conclusion that, while in
the absence of traffic congestion a star-like (centralized) topology
is the optimal one (as briefly discussed in the introduction) when the
density of information-packets traveling through the net is above a
given threshold (i.e. when there is {\it flow congestion}) the optimal
topology is a highly homogeneous one. In this last, all nodes have
essentially the same degree, the same betweenness, and short loops are
absent (see figure 1 in \cite{Catalans}). These networks resemble
enormously the entangled and Ramanujan nets described above.

For a similar comparison between centralized and de-centralized
transport in networks, see \cite{Congestion}. Also here,
decentralized, highly homogeneous, structures emerge as optimal ones
under certain circumstances.

\subsection{Network structures from selection principles}

In a recent letter, Colizza et al. \cite{Maritan} have addressed the
search of optimal topologies by using different {\it selection
principles}. In particular they minimize a global cost function
defined as
\begin{equation}
 H_{\alpha} = \sum_{i < j} d_{ij} (\alpha)
\end{equation}
with
\begin{equation}
 d_{ij} (\alpha) = \min_P \sum_{p\in P: i \rightarrow j} k_p^{\alpha}. 
\end{equation}
This is, the cost function is the sum over pairs of sites of their
relative $\alpha$-dependent distance. The distance between two nodes
is the minimum of the sum of $k_p^\alpha$ ($k_p$ is the degree of node
$p$) over all possible paths connecting the two nodes. In this way,
for $\alpha=0$ the standard distance is recovered. On the other hand,
for large values of $\alpha$, highly connected vertices (hubs) are
strongly penalized. In this way, and rephrasing the authors of
\cite{Maritan}, the generalized definition of the distance ``captures
the conflict between two different trends: the avoidance of long paths
and the desire to skip heavy traffic''. For small values of $\alpha$
(standard distance) the best topology include central hubs
(centralized communication), while for $\alpha >1$ the dominant
tendency is towards degree minimization, leading to open tree-like
structures. For intermediate cases, i. e. $\alpha=0.5$ topologies
extremely similar to entangled networks, with long loops and very high
homogeneity, have been reported to emerge (see figure 3c in
\cite{Maritan}).

\subsection{Performance of neural networks}

Recently Kim concluded that neural networks with a clustering
coefficient as small as possible exhibit much better performance than
others \cite{BJK}. Entangled nets have a very low clustering
coefficient as only large loops exist and, therefore, they are natural
candidates to constitute an excellent topology to achieve good
performance and high capacity in artificial neural networks.

\subsection{Robustness and resilience}

The problem of constructing networks whose robustness against random
and/or intentional removal is as high as possible has attracted a lot
of attention. In particular, in a recent paper \cite{2peak}, it has
been shown that for generalized random graphs in the limit
$N\to\infty$ the most robust topology (in the sense that the
corresponding percolation threshold is as large as possible) has a
degree distribution with no more than $3$ distinct node
connectivities; i.e. with a homogeneous degree-distribution as
homogeneous as possible.

To study the possible connection with our extremely homogeneous
entangled networks, let us recall that the topology we have considered
to initialize the improved version of the optimization algorithm
(i.e. random $k$-regular graphs) is already the optimal solution for
robustness-optimization against (combined) errors and attacks in
random networks \cite{2peak}. A natural question to ask is whether
further $Q$-optimization has some effect on the network
robustness. This and related questions have been analyzed in
\cite{Entangled}, where it was shown that indeed, $Q$ minimization
implies a robustness improvement. This occurs owing to the generation
of non-trivial correlations in entangled structures, which place them
away from the range of applicability of the result in \cite{2peak}
(they are not random networks).  Hence, entangled networks are also
extremely efficient from the robustness point of view. This conclusion
also holds for {\it reliability} against link removal \cite{Myrvold}.

\section{VII. The connection with the Bethe lattice}

In this section we will show how entangled networks are related to the
infinite regular tree called {\it Bethe lattice} (or Bethe tree) in
the physics literature. 

Intuitively, whenever around a given node of a regular graph only
large loops are present, the neighborhood of such node looks as the
one in a Bethe tree, up to a distance equal to the half of the
shortest loop length (see for instance the center of the graph
depicted to the right of figure \ref{NoCages}).  Therefore, if the
girth diverges in a family of graphs, the neighborhood of each node
tends to look locally like a Bethe, up to a growing distance.


We notice that the asymptotic spectral gap of regular graphs is
bounded by the one of the Bethe tree, $\lambda_2 \leq k- 2
\sqrt{k-1})$ (see equation (\ref{inequality2})) \cite{bethegap}. 
We can express this bound in terms of the spectral properties of the
adjacency matrix (since the graphs are regular, the results can be
easily translated to $L$ and $\LN$).

The moments $m_n$ of the adjacency matrix eigenvalues $\mu_i$ can be
trivially expressed through the trace of the $n$-th power of $A$ \cite{LPS}:
\begin{equation}
  \label{eq:moms} m_n = \frac1N \sum_i \mu_i^n = \frac1N \sum_i
  (A^n)_{ii}.
\end{equation}
It is easy to see that the diagonal elements of $A^n$ represent the
number of paths of length $n$ starting and ending at the corresponding
node. Clearly, if $n$ is smaller than the girth of the graph the paths
do not contain loops and their number is equal to the number of such
paths for a node of the Bethe lattice and therefore equal for all the
nodes. The eigenvalue moments for the infinite Bethe lattice can be
obtained in a similar way, with the only difference that the
eigenvalue distribution is now a continuous spectral density
\cite{density}. The sum in equation~(\ref{eq:moms}) is now an integral
and the adjacency matrix is a linear operator; however it is still
possible to express the $n$-th moments as the number of length-$n$
paths from a node to itself. As a result the moments $m_n$ of the
adjacency matrix eigenvalues of a regular graph with girth $g$ are
equal to the moment of the spectral density 
\begin{equation}
  \label{eq:specden} \rho(\mu) = \frac{k \sqrt{4(k-1)-\mu^2}}{2 \pi
  (k^2 -\mu^2)}\ , \quad \text{for } |\mu| \le 2\sqrt{k-1}
\end{equation}
of the Bethe tree \cite{bethedensity}, for every $n<g$. 

The previous argument can be generalized and it can be rigorously
proven \cite{mckay} that for an infinite sequence $G_n$ of $k$-regular
graphs, if the number of loops of length $l$ grows slower than the
number of nodes for every $l$ then, the spectral density tends to the
one in equation~(\ref{eq:specden}). Hence, we expect a fast
convergence for graphs with large girth as the ones discussed before.

\begin{figure}
\centerline{ 
\psfig{file=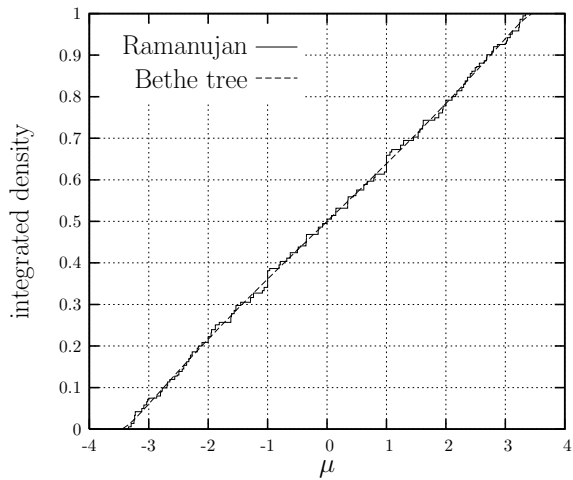,width=7.5cm}}
\caption{Integrated spectral density as a function of the adjacency matrix eigenvalues 
for the Bethe lattice (infinite size) and for a Ramanujan graph with
$N=6840$, both with $k=4$.}
\label{spden}
\end{figure}

To illustrate this, in figure \ref{spden} we plot the Bethe lattice
integrated spectral density as a function of the eigenvalues for $k=4$
and compare it with the one calculated for a Ramanujan graph with
$6840$ nodes ($q=19$) and also $k=4$ ($p=3$). The agreement is very
remarkable given the finite size of the Ramanujan graph.

Therefore, both from the (local) topological and the spectral point of
view, a family of entangled networks (or Ramanujan graphs) with
growing size and girth can be used as the best possible way to
approach infinite Bethe lattices with a sequence of finite lattices.
This would have applications in a vast number of problems in physics,
for which exact solutions in the Bethe lattice exist, but comparisons
with numerics in sufficiently large lattices are difficult to obtain.
In particular, approximations of Bethe lattice obtained by truncating
the number of generations (Cayley trees) include a large (extensive)
amount of boundary effects (losing the original homogeneity) and
are therefore not a convenient finite approximation. Good
approximations should avoid strong boundary effects and preserve the
Bethe lattice homogeneity, and entangled networks can play such a
role.


\section{VIII. A way out of homogeneity: weighted dynamics}

While most of the literature on synchronization refers to Laplacian
couplings as specified by equation (\ref{pecora}), different coupling
functions can be relevant in some contexts. For example, another
natural choice would be
\begin{equation}
\dot{x}_i = 
F(x_i) + {\sig\over k_i}   \sum_{j\in n.n. i} [ H(x_j) -H(x_i)] =
F(x_i) - {\sig\over k_i}   \sum_j  L_{ij} H(x_j),
\label{norma}
\end{equation}
relevant in cases where the joint effect of the $k_i$ neighbors of
node $i$ is normalized by the connectivity $k_i$. With this type of
dynamics, the effect of neighbors has the same weight for all nodes,
while in the absence of the normalization factor, $k_i$, sites with
higher connectivity are more strongly coupled to their neighbors than
loosely connected ones. This describes properly real-world situations,
as for instance neural networks, where the influence of the
neighboring environment on the node dynamics does not grow with the
number of connections. Observe that, even if the underlying topology
is unweighted and undirected, the normalization factor is such that
{\it the dynamics is directed and weighted}, owing to the presence of
$k_i$ in eq.(\ref{norma}).

In a recent paper Motter and coauthors have undertaken a study of
synchronizability by using a generalization of the previous normalized
dynamics eq.(\ref{norma}), by employing a normalization factor
$k_i^\beta$, with $\beta \geq 0$ \cite{Zhou}. These authors conclude
that the most robust network synchronizability is obtained for
$\beta=1$, which leads back to Eq.(\ref{norma}).

{\it What is the influence of the normalization factor on the results
reported on this paper?}

First of all, the optimal synchronizability problem, as discussed in
section III, can be straightforwardly translated for the new dynamics
just by replacing Eq.(\ref{pecora}) by Eq.(\ref{norma}). The optimal
topology for synchronizability in ``normalized'' dynamical processes
is that minimizing the eigenratio $Q_{norm}=\lambda'_N/\lambda'_2$,
($\lambda'_i$ denote the normalized Laplacian eigenvalues).  Hence, by
employing the normalized dynamics, we have a new class of optimization
problems, analogous but different to the ones studied before along
this paper.

\begin{figure}
\centerline{
\psfig{file=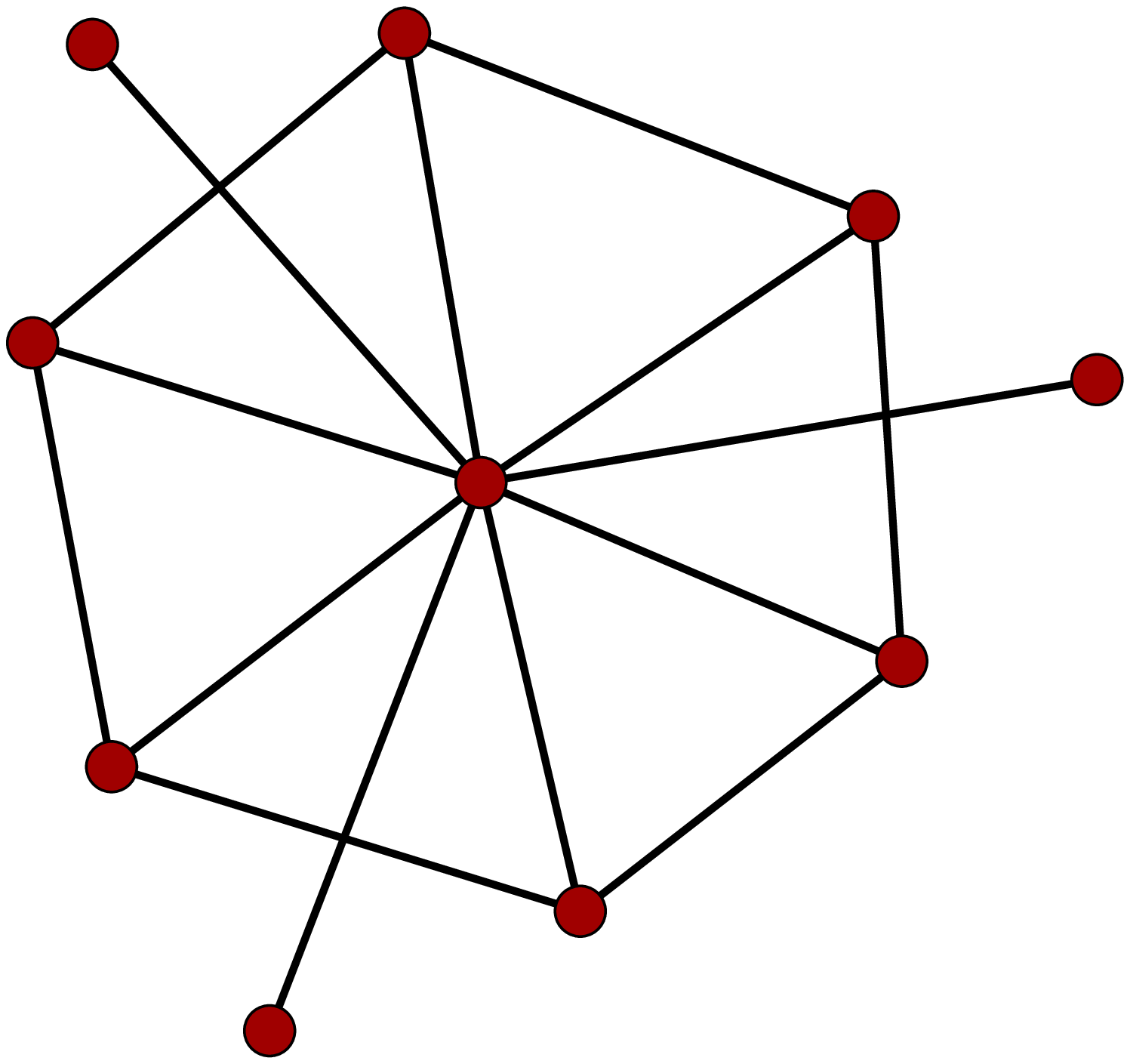,width=4.0cm}
\psfig{file=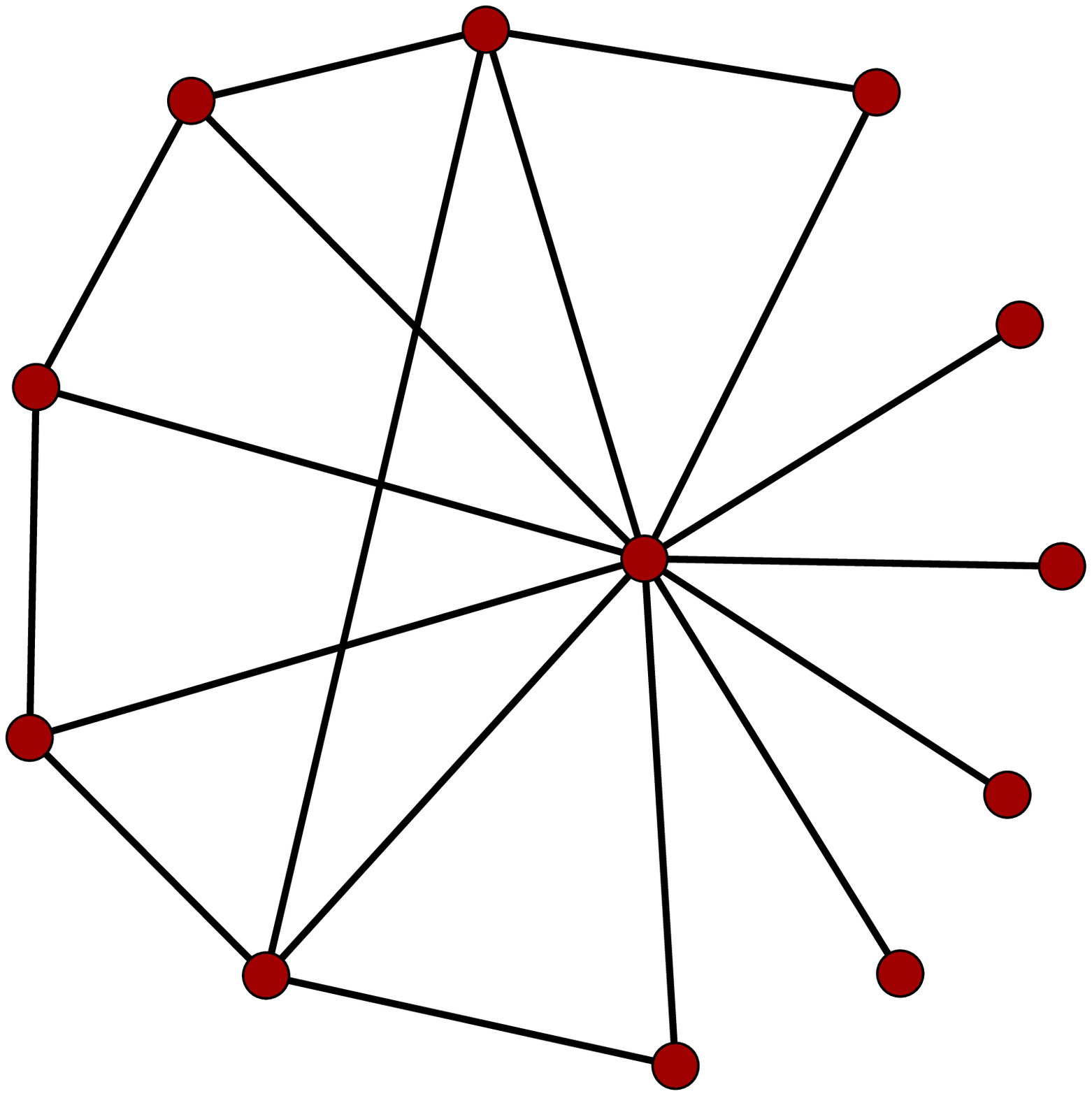,width=4.0cm}
\psfig{file=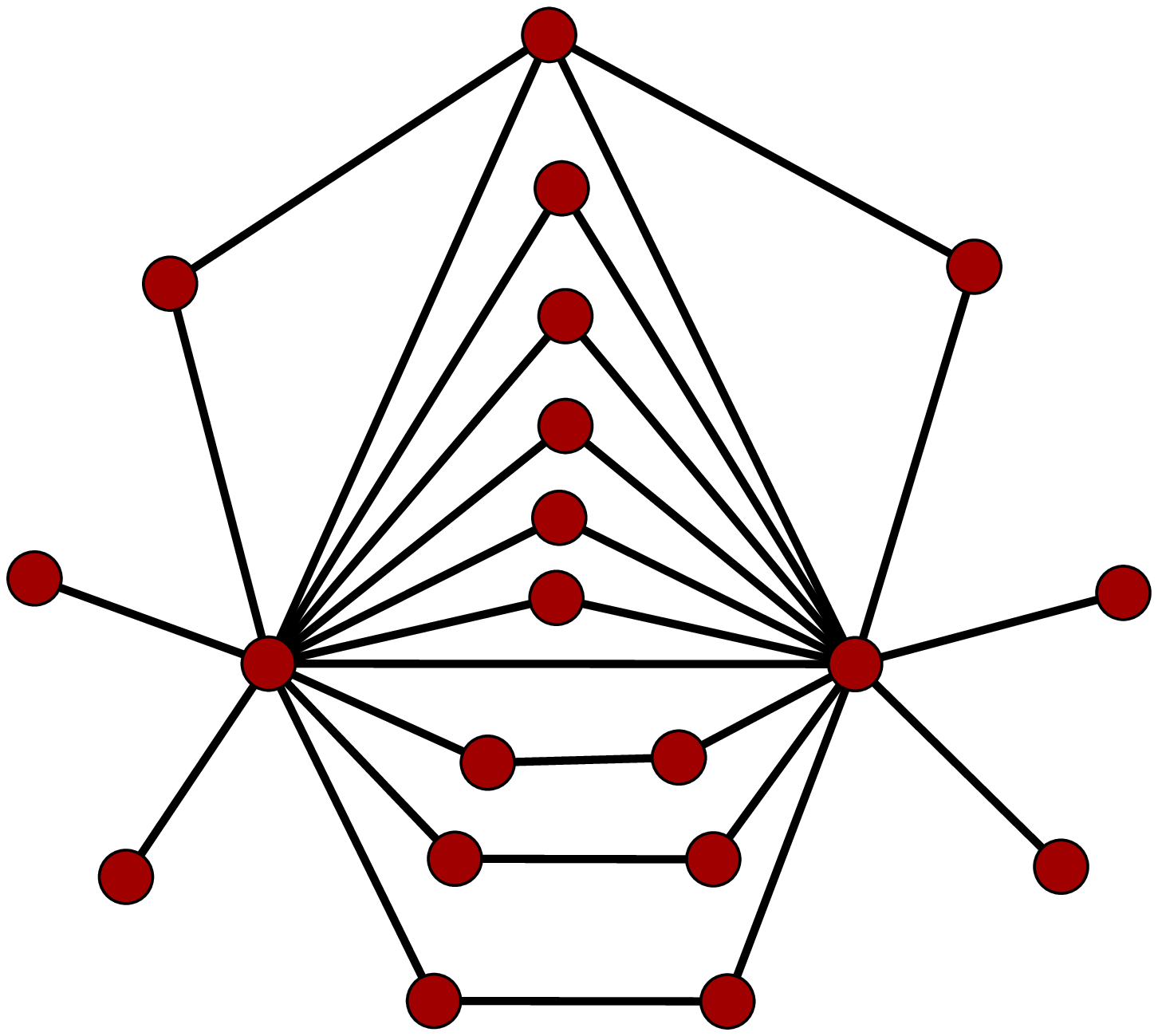,width=4.0cm}}
\caption{Optimal nets for normalized-Laplacian dynamics, 
with $\langle k \rangle=3$ and $N=10, 12$, and $20$ respectively.}
\label{newgraphs}
\end{figure}

\begin{figure}
\centerline{
\psfig{file=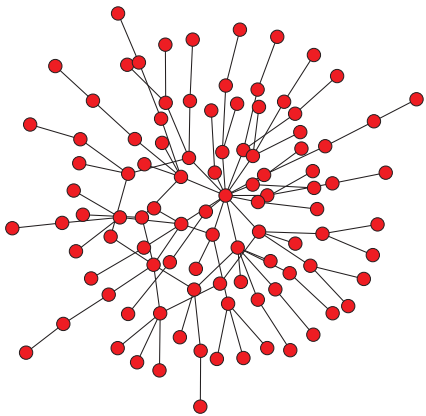,width=6.0cm}}
\caption{Optimal net for normalized-Laplacian dynamics, 
with $\langle k \rangle=3$ and $N=100$.}
\label{newgraphs2}
\end{figure}

We have implemented different versions of our modified simulated
annealing algorithm (in its more general, not degree-conserving,
version) to optimize either $Q_{norm}$ or the normalized spectral gap
$\lambda'_2$. 
After going through the optimization algorithm with
normalized-Laplacian eigenvalues, as described in section V, one
observes that the emerging optimally synchronizable nets have a {\it
non-homogeneous structure}. For some relatively small values of $N$
and $\langle k \rangle=3$ the optimal topologies are shown in figure
\ref{newgraphs}, while in figure \ref{newgraphs2} we depict an optimal
net with $100$ nodes.

\begin{figure}[h]
\centerline{
\psfig{file=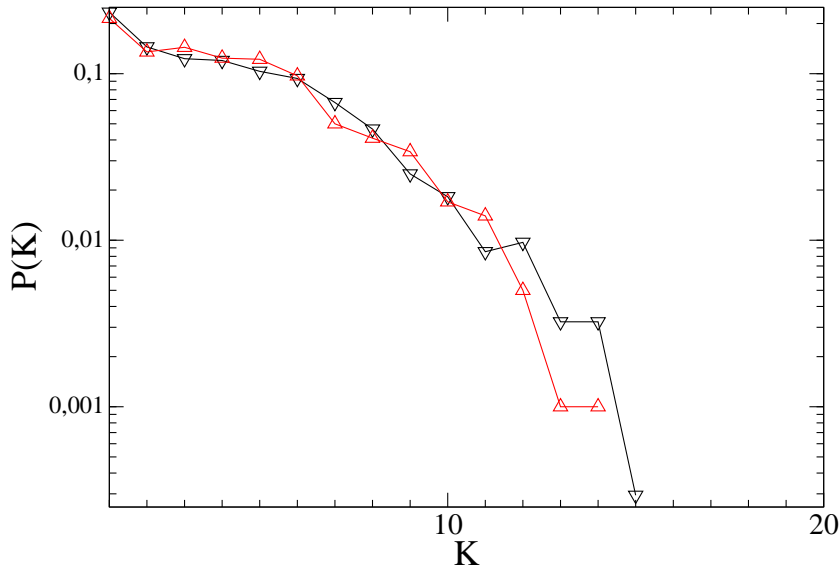,width=9.5cm,angle=-90}}
\caption{Semi-logarithmic plot of the degree distribution of optimal weighted
networks with $\langle k \rangle =4$ and (i) $N=200$ averaged over
$17$ runs (black curve with down-triangles) and (ii) $N=1000$ for a
single run (red curve with up-triangles).}
\label{degree}
\end{figure}

Observe the remarkable difference with previously studied networks:
here homogeneity is drastically reduced. To illustrate this, in figure
\ref{degree} we plot the degree distribution obtained when optimizing
for $\langle k \rangle=4$ with $N=200$ and $N=1000$ respectively. The
resemblance between these two distributions seems to indicate that a
limit degree-distribution exists. Roughly speaking it decays faster
than exponentially, but still with a much higher degree of
heterogeneity than before where we had delta-peaked distributions.

The new emerging topologies exhibit a competition between the
existence of central nodes and peripheral ones similarly to the
networks studied in \cite{Congestion}. By maximizing the spectral gap
$\lambda'_2$, rather than $\lambda'_N/\lambda'_2$ we obtain very
similar results (not shown here). These new optimal topologies are
very similar to those obtained by Colizza et al. \cite{Maritan} in
cases where the minimization of node-degree dominates (see figure 3d
in
\cite{Maritan}).

Let us remark that for the normalized-Laplacian, some eigenvalues
bounds, analogous to those for $L$ exist; in particular \cite{Chung}
\begin{equation}
0 < \lambda'_2 \leq {N \over N-1} \leq \lambda'_N \leq 2.
\label{nl}
\end{equation}
However, concepts analogous to ``expanders'' or ``Ramanujan'' have not
been defined for general, non-regular, graphs.

Finally, it is interesting to observe that our results are in apparent
contradiction with the conclusion by Motter et al. \cite{Zhou} that
for their case $\beta=1$, corresponding to our normalized Laplacian,
and large {\it sufficiently random networks} the eigenratio $Q$ does
not depend on details of the network topology, but only on its mean
degree \cite{Zhou}. If this was indeed the case for any network
topology, our optimization procedure would be pointless. Instead,
applying the minimization algorithm, starting from any arbitrary
finite random network, we observe a progressive $Q$ optimization, and
highly non-trivial non-random optimal structures are actually
generated. This apparent contradiction is likely to be due to the
building up of non-trivial correlations during the optimization
process, which converts our network into highly non-random structures
far away from the {\it sufficiently random} requirement in
\cite{Zhou}.

\section{IX. Discussion and Conclusions}

In this paper we have reviewed recent developments in the design of
optimal network topologies.

First of all we have related the problem of finding optimal topologies
for many dynamical and physical problems, such as optimal
synchronizability and random walk flow on networks, to the search of
networks with large spectral gaps.  This connection allows us to
relate optimal networks to expanders and Ramanujan graphs, well-known
concepts in graph-theory which, by construction, have large spectral
gaps. In one of the sections we have given, following the existing
mathematical literature, a recipe to explicitly construct Ramanujan
graphs.

On the other hand, we have employed a simulated annealing algorithm
which selects progressively networks with better synchronizability,
i.e. networks with larger spectral gaps. When applied to problems with
un-normalized Laplacian dynamics this algorithm leads to what we call
``entangled'' topologies.  These are optimal expanders, and can be
loosely described as being extremely homogeneous, having long loops,
poor modularity, and short node-to-node distances. They are optimal or
almost optimal for many different communication and flow processes
defined on regular networks. In particular these topologies are
relevant for the design of efficient communication networks,
construction of error-correcting codes with very efficient encoding
and decoding algorithms, de-randomization of random algorithms,
traffic problems with congestion, and analysis of algorithms in
computational group theory.  Remarkably, they also provide a good
finite-size approximation to Bethe lattices.

Even though these topologies play an important role in human-designed
networks, especially in Computer Science and Algorithmics, they do not
seem to appear frequently in Nature. Indeed, most of the topologies
described in recent years for biological, ecological, social, or
technological networks exhibit a very heterogeneous scale-free degree
distribution \cite{Strogatz,Laszlo,Newman,Porto,AleRomu,Boccaletti},
at odd with the extremely homogeneous entangled topologies.  To
justify this, first of all we should emphasize that optimal entangled
networks are obtained performing a {\it global} optimization
procedure, which might be not very realistic. For example, by adding a
new node to a given optimal network, in some cases the full entangled
topology has to be altered to convey with the requirement of
maximizing the spectral gap. This is certainly impractical for growing
networks.

On the other hand, we should also note that there are ingredients that
could be (and actually are) relevant for the determination of optimal
networks in real-world problems and that have not being considered
here. A non exhaustive list of examples is:

\begin{itemize}
\item nodes might be non equivalent,

\item links could support different variable weights,

\item directed networks might be mandatory in some cases,

\item in some real-world networks nodes are embedded into a geography 
and, therefore, the distances between them should be taken into
account as a sort of quenched disorder (see for instance,
\cite{Gastner,Barthelemy} for application to communication infrastructures).

\end{itemize}

However, our entangled networks are still relevant for any type of
un-directed un-weighted networks in which none of these effects plays
an important role, and where communication properties are to be
maximized. Numerous examples have being cited above and along the
paper.

In the last section of the paper, we have ``enlarged our horizon'' by
studying a different type of dynamical processes, that could be argued
to be more realistic in some circumstances: instead of analyzing
Laplacian coupling between the network nodes we shifted to the study
of normalized-Laplacian coupling, which lead to weighted and directed
dynamics. This is characterized by the presence of a ``normalized
dynamics'' in which the relative influence of the neighbors on a given
site does not depend on the site connectivity. In particular, this
type of dynamics might be more adequate to describe random walks and
general diffusion processes on general (not necessarily regular)
networks. In this case, an optimization procedure analogous to the one
described before, leads to much more heterogeneous optimal topologies,
including hubs, and a much broader degree distribution. This makes the
emerging optimal networks closer to real-world topologies than
entangled networks, while still keeping the global-optimization
perspective. Along this same line of reasoning, other alternative
types of dynamics could be introduced, as for instance the
``load-weighted'' ones first proposed in
\cite{Chavez,Zhou} as a generalization of the normalized-Laplacian
dynamics. A very interesting, open question, is to define optimization
processes along the lines of this paper, leading to scale free
topologies. 

We hope this work will encourage further research on this fascinating
question of optimal network topologies, their evolution, their
application to human designed networks and their connection with
real-world complex networks.

\vspace{1cm}
 
\begin{acknowledgments} We acknowledge useful discussions with
 D. Cassi with whom we are presently exploring the connection with the
 Bethe lattice. We are especially thankful to P. I. Hurtado, our
 coauthor in the paper where entangled networks were first introduced,
 for a very enjoyable collaboration and a critical reading of the
 manuscript. F. N. acknowledges the kind hospitality in Granada where
 this work was initiated. Finally, we acknowledge financial support
 from the Spanish MEyC-FEDER, project FIS2005-00791, and from Junta de
 Andaluc{\'\i}a as group FQM-165.
\end{acknowledgments}



\begin{thebibliography}{99}


\bibitem{synchro}
     T. Nishikawa, A. E. Motter, Y.-C. Lai, and F. C. Hoppensteadt,
     Phys. Rev. Lett. {\bf 91}, 014101 (2003).
     H. Hong, B. J. Kim, M. Y. Choi, and H. Park
      Phys. Rev. E. {\bf 69}, 067105 (2004). H. Hong, M. Y. Choi, and
      B. J. Kim, Phys. Rev. E {\bf 65}, 026139 (2002); Phys. Rev. E
      {\bf 65}, 047104 (2002).


\bibitem{Torres} J. J. Torres, M. A. Mu\~noz, J. Marro, and P. L. Garrido,
Neurocomputing, {\bf 58-60}, 229 (2004); and references therein.


\bibitem{BJK} B. J. Kim, 
Phys. Rev. E {\bf 69}, 045101(R) (2004).  See also, P. McGraw and
M. Menzinger, Phys. Rev. E {\bf 72}, 015101(R) (2005).


\bibitem{GG} G. Grinstein and R. Linsker, 
 Proc. Natl. Acad. Sci. USA, {\bf 102}, 9948 (2005).



\bibitem{Catalans}
     R. Guimera, A. Arenas, A. Diaz-Guilera, F. Vega-Redondo,
     A. Cabrales,
      Phys. Rev. Lett. {\bf 89}, 248701 (2002).



\bibitem{Congestion} D. J. Ashton, T. C. Jarrett, and N. F. Johnson,
Phys. Rev. Lett. {\bf 94}, 058701 (2005).

\bibitem{Maritan} 
V. Colizza, J. R. Banavar, A. Maritan, and A. Rinaldo,
Phys. Rev. Lett. {\bf 92}, 198701 (2004).



\bibitem{Lovasz} L. Lov\'asz, 
in \emph{Combinatorics, Paul Erd\"os is Eighty} (V2), Keszthely,
Hungary, pp. 1-46, (1993).



\bibitem{dani} E. M. Bollt and D. ben-Avraham,
New Journal of Phys. {\bf 7}, 26 (2005). 

\bibitem{Strogatz} S. H. Strogatz, Nature {\bf 410}, 268 (2001).

\bibitem{Laszlo} A. L. Barab\'asi,  
Rev. Mod. Phys. {\bf 74}, 47 (2002).

 \bibitem{Newman} M. E. J. Newman,
SIAM Review {\bf 45}, 167 (2003).

\bibitem{Porto} S. N. Dorogovtsev and J. F. F. Mendes, {\it Evolution of
    Networks: From Biological Nets to the Internet and WWW}, Oxford
  University Press, Oxford (2003).
  
\bibitem{AleRomu} R. Pastor Satorras and A. Vespignani, {\it Evolution and
    Structure of the Internet: A Statistical Physics approach}, Cambridge
  University Press (2004).

\bibitem{Boccaletti} S. Boccaletti, V. Latora, Y. Moreno, 
M. Chavez, and D.-U. Hwang, Phys. Rep. {\bf 424}, 175 (2006).


\bibitem{Chung}
 F. Chung, {\it Spectral Graph Theory},
Number 92 in CBMS Regional Conference Series in Mathematics. Am. Math.
Soc., 1997. See also, F. Chung, L. Lu, and V. Vu,
Internet Mathematics, {\bf I}, 257 (2004).

\bibitem{Bollobas} 
B.  Bollob\'as, {\it Extremal Graph Theory} Academic Press, New
York. 1978. W. Tutte, {\it Graph Theory As I Have Known It}, Oxford
U. Press, New York, (1998).  

\bibitem{GN} M. Girvan, M. E. J. Newman, 
Proc. Natl. Acad. Sci. USA {\bf 99}, 7821-7826 (2002).
M. E. J. Newman, M. Girvan, Phys. Rev. E {\bf 69}, 026113 (2004).
See also, L. Donetti and M. A. Mu\~noz, 
J. Stat. Mech.: Theor. Exp. (2004) P10012;
in ``{\it Modeling Cooperative behavior in the social sciences}'', AIP
Conf. Proc. 779, 104 (2005); arXchic:Physics/0504059.


\bibitem{Sarnak} P. Sarnak,
Notices Amer. Math. Soc. {\bf 51}, 762 (2004). 

\bibitem{Valette} G. Davidoff, P. Sarnak, and A. Valette, {\it Elementary Number Theory, 
Group Theory and Ramanujan Graphs}, London Math. Soc. Students Texts,
55, Cambridge (2003).

\bibitem{AB} The original statement of the lower bound
 is due to N. Alon and R. Boppana, and appears in A. Nilli,
Discrete Math., {\bf 91}, 207 (1991).


\bibitem{Entangled} L. Donetti, P. I. Hurtado and  M. A. Mu\~noz,
Phys. Rev. Lett.  {\bf 95}, 188701 (2005).


\bibitem{Pecora}  
M. Barahona and L. M. Pecora,
Phys. Rev. Lett. {\bf 89}, 054101 (2002).  See also, L. M. Pecora and T. L.
Carroll, Phys. Rev. Lett. {\bf 64}, 821 (1990); ibid, {\bf 80}, 2109
(1998).
L. M. Pecora, Phys. Rev. E {\bf 58} 347 (1998).
L. M. Pecora and T. L. Carroll, Phys. Rev. Lett. {\bf 80}, 2109
(1998).




\bibitem{Wang}  X. F. Wang and G. Chen,
Int. J. Bifurcation Chaos Appl. Sci. Eng. {\bf 12}, 187
(2002). X. F. Wang and G. Chen, IEEE Trans. Circuits and Systems I
{\bf 49}, 54 (2002).  

 
\bibitem{Mohar} B. Mohar, 
in {\it Graph Theory, Combinatorics, and Applications, Vol 2,}
Ed. Y. Alavi, G. Chartrand, O. R. Oellermann, and A. J Schwenk,
Wiley, New York, 1991. pp. 871.
 


\bibitem{Nielsen}  
 $http://www.math.ias.edu/~boaz/ExpanderCourse/$.
See also, $http://www.qinfo.org/people/nielsen/blog/?p=222$


\bibitem{PL} P. Pons and M. Latapy, Preprint. arXiv:physics/0512106. 


\bibitem{LPS} A. Lubotzky, R. Phillips, and P. Sarnak, 
Combinatorica, {\bf 8}, 261 (1988).


\bibitem{Margulis} G. A. Margulis, Prob. of Info. Trans. 
{\bf 9}, 325 (1975). See also, O. Reingold, S. Vadhan, and
A. Wigderson, Ann. of Math. {\bf 155}, 157 (2002).





\bibitem{Chiu} P. Chui, Combinatorica, {\bf 12}, 275 (1992).


\bibitem{Code} Our computer code to generate Ramanujan graphs is available 
upon request to any of the authors.




\bibitem{betweenness} 
The betweenness centrality is defined as the average number of
shortest paths, connecting every possible couple of nodes, passing
through each site. An edge-betweenness can be analogously defined for
the number of shortest paths passing through each link.

\bibitem{SA} S. Kirkpatrick, C. D. Gelatt, and M. P. Vecchi, 
Science {\bf 220} 671 (1983).  N. Metropolis, A. W. Rosenbluth,
M. N. Rosenbluth, A. H. Teller, and E. Teller, J. Chem. Phys. {\bf 21}
1087 (1953).


\bibitem{Penna} T. J. P. Penna, Phys. Rev. E {\bf 51}, R1 (1995).


\bibitem{Cages}
Eric W. Weisstein. "Cage Graph." From MathWorld--A Wolfram Web
Resource. $http://mathworld.wolfram.com/CageGraph.html$; and
references therein. See also, R. C. Read and R. J. Wilson,{\it An
Atlas of Graphs}, Oxford, England: Oxford University Press, pp. 263
and 271-274, 1998.


\bibitem{2peak} A.X.C.N. Valente, A. Sarkar, and H.A. Stone,
Phys. Rev. Lett. {\bf 92}, 118702 (2004).

\bibitem{Myrvold}
     W. Myrvold,
     {\it Proceedings of the $8th$ Int. Conf. on Graph Theory,
       Combinatorics, Algorithms, and Applications}, Vol. II, pag. 650
     (1998).

\bibitem{bethegap} 
See D. Cassi, Phys. Rev. B {\bf 45}, 454 (1992); and references
therein.

\bibitem{density} Note that as the tree is infinite, its spectrum is continuous, and
therefore, it is more meaningful to refer to ``spectral densities''.

\bibitem{bethedensity} H. Kesten, \textit{Trans. AMS} \textbf{92}, 336
  (1959).



\bibitem{mckay} B. D. McKay, \textit{Linear Algebra and its
    applications} \textbf{40}, 203 (1981)

\bibitem{Future} D. Cassi, F. Neri, L. Donetti, and M.A. Mu\~noz;
preprint 2006.

\bibitem{Zhou}  A. E. Motter, C. Zhou, and J. Kurths,
Phys. Rev E {\bf 71}, 016116 (2005); AIP Conference Proceedings
{\bf 776}, 201 (2005); EuroPhys. Lett. {\bf 69}, 334 (2005).

\bibitem{Gastner} M. T. Gastner and M. E. J. Newman,
arXiv:cond-mat/0603278.

\bibitem{Barthelemy}  M. Barthelemy and A. Flammini,
arXiv:physics/0601203.
  
\bibitem{Chavez} M. Chavez, D.-U. Hwang, A. Amann, 
H. G. E. Hentschel, and S. Boccaletti, Phys. Rev. Lett. {\bf 94},
218701 (2005).  

\end{thebibliography}
\end{document}